\documentclass[11pt,a4paper]{article}
\usepackage{epstopdf}
\usepackage{graphicx}
\usepackage{authblk}
\usepackage[hyperindex,breaklinks]{hyperref}

\title{Simulation of the CMS Resistive Plate Chambers}

\author{R. Hadjiiska$^a$\thanks{Corresponding author: e-mail: roumyana.mileva.hadjiiska@cern.ch}, L. Litov$^a$, B. Pavlov$^a$, P. Petkov$^a$, A. Dimitrov$^a$, K. Beernaert$^b$, A. Cimmino$^b$, S. Costantini$^b$, G. Garcia$^b$, J. Lellouch$^b$,
A. Marinov$^b$, A. Ocampo$^b$, N. Strobbe$^b$, F. Thyssen$^b$, M. Tytgat$^b$, P. Verwilligen$^b$\thanks{Now at Universita e INFN, Sezione di Bari.},
E. Yazgan$^b$, N. Zaganidis$^b$, A. Aleksandrov$^c$, V. Genchev$^c$, P. Iaydjiev$^c$, M. Rodozov$^c$, M. Shopova$^c$, G. Sultanov$^c$, Y. Ban{$^d$}, J. Cai{$^d$}, Z. Xue{$^d$}, Y. Ge{$^d$}, Q. Li{$^d$}, S. Qian{$^d$}, C. Avila{$^e$}, L.F. Chaparro{$^e$}, J.P. Gomez{$^e$}, B. Gomez Moreno{$^e$}, A.F. Osorio Oliveros{$^e$}, J.C. Sanabria{$^e$}, Y. Assran{$^f$}, A. Sharma{$^g$}, M. Abbrescia{$^h$}, A. Colaleo{$^h$}, G. Pugliese{$^h$}, F. Loddo{$^h$}, C. Calabria{$^h$}, M. Maggi{$^h$}, L. Benussi{$^i$}, S. Bianco{$^i$}, S. Colafranceschi{$^i$}, D. Piccolo{$^i$}, C. Carrillo{$^j$}, O. Iorio{$^j$}, S. Buontempo{$^j$}, P. Paolucci{$^j$}, P. Vitulo{$^k$}, U. Berzano{$^k$}, M. Gabusi{$^k$}, M. Kang{$^l$},  K.S. Lee{$^l$}, S.K. Park{$^l$}, S. Shin{$^l$}, M.S. Kim{$^m$}, H. Seo{$^m$}, J. Goh{$^m$}, Y. Choi{$^m$} and M. Shoaib{$^n$}\\
\footnotesize
\textit{$^a$}University of Sofia, Faculty of Physics, Atomic Physics Department,\\
  5, James Bourchier Boulevard, BG-1164 Sofia, Bulgaria\\
\footnotesize
\textit{$^b$}Ghent University, Department of Physics and Astronomy,\\
  Proeftuinstraat 86, B-9000 Gent, Belgium\\
\footnotesize
\textit{$^c$}Bulgarian Academy of Sciences,\\
  Inst. for Nucl. Res. and Nucl. Energy,\\
  Tzarigradsko shaussee Boulevard 72, BG-1784 Sofia, Bulgaria\\
\footnotesize
\textit{$^d$}Peking University, School of Physics,\\
  CN-100871 Beijing, China\\
\footnotesize
\textit{$^e$}Universidad de Los Andes,\\
  Apartado A\'ereo 4976, Carrera 1E, no. 18A 10, CO-Bogot\'a, Colombia\\
\footnotesize
\textit{$^f$}Academy of Scientific Research and Technology of the Arab Republic of Egypt,\\
  101 Sharia Kasr El-Ain, Cairo, Egypt\\
\footnotesize
\textit{$^g$}Panjab University, Department of Physics,\\
  Chandigarh Mandir 160 014, India\\
\footnotesize
\textit{$^h$}	Universita e INFN, Sezione di Bari,\\
  Via Orabona 4, IT-70126 Bari, Italy\\
\footnotesize
\textit{$^i$}INFN, Laboratori Nazionali di Frascati (LNF),\\
  PO Box 13, Via Enrico Fermi 40, IT-00044 Frascati, Italy\\
\footnotesize
\textit{$^j$}Universita e INFN, Sezione di Napoli,\\
  Complesso Univ. Monte S. Angelo, Via Cintia, IT-80126 Napoli, Italy\\
\footnotesize
\textit{$^k$}Universita e INFN, Sezione di Pavia,\\
  Via Bassi 6, IT-Pavia, Italy\\
\footnotesize
\textit{$^l$}Korea University, Physics Department,\\
  Seoul Cheongryangri 143-701, Republic of Korea\\
\footnotesize
\textit{$^m$}Sungkyunkwan University, Department of Physics,\\
  2066, Seobu-ro, Jangan-gu, Suwon, Gyeonggi-do, Republic of Korea\\
\footnotesize
\textit{$^n$}National Centre for Physics, Q.A.U Campus, \\
  Shahdra Valley Road, Islamabad 44000, Pakistan\\
}

\date{}

\begin{document}

\maketitle

\begin{abstract}The Resistive Plate Chamber (RPC) muon subsystem contributes significantly to the formation of the trigger decision and reconstruction of the muon trajectory parameters. Simulation of the RPC response is a crucial part of the entire CMS Monte Carlo software and directly influences the final physical results. An algorithm based on the parametrization of RPC efficiency, noise, cluster size and timing for every strip has been developed. Experimental data obtained from cosmic and proton-proton collisions at $\sqrt{s}=7$ TeV have been used for determination of the parameters. A dedicated validation procedure has been developed. A good agreement between the simulated and experimental data has been achieved.


\end{abstract}

\section{The CMS Muon system}

\par The CMS (Compact Muon Solenoid) is one of the two general-purpose detectors located at the LHC (Large Hadron Collider). The main part of the detector is a superconducting solenoid providing axial magnetic field of $3.8$ T inside the coil and $1.8$ T in the return yoke. 
The silicon tracker system, the crystal electromagnetic calorimeter, and the sampling brass/scintillator hadronic calorimeter are enclosed within the coil. The muon system is deployed outside of the solenoid and is situated between the layers of the steel return yoke.

\par Three different technologies have been used to trigger and measure muons: drift tubes (DT) (pseudo-rapidity $\left|\eta\right| < 1.2$); cathode strip chambers (CSC) ($0.9 < \left|\eta\right| < 2.4$) and resistive plate chambers (RPC) ($\left|\eta\right| < 1.6$). The muon system consists of four stations in the barrel and in the endcaps. In the barrel part, each station consists of 12 DT layers packed on both sides by two RPC in the first and the second stations and one RPC in the third. The outermost DT station has eight layers and one RPC. In the endcaps, each muon station consists of six CSC detection planes complemented by the RPC in the three inner stations. A detailed description of the CMS detector can be found elsewhere \cite{cms}.

\par The RPC system consists of 480 barrel chambers and 432 end-cap chambers and covers an active area of $2953$ m$^2$ \cite{CAR}. This makes the RPC system the largest sub-detector at CMS.

\par The ability to work at a high rate of ionizing particles up to $1$ {kHz/cm$^2$} has been ensured by Bakelite double-gap RPCs operating in avalanche mode. A double-gap encompasses two single gaps mounted one on the top of each other with common readout strips situated in between them. The rate limitations are due to the time needed to re-establish the electric field after an avalanche in a gas gap. The voltage drop is proportional to the average avalanche charge in the gas gap and the electrode resistivity. The analytical description of the rate effects in RPC is based on simple DC model (see \cite{Die}, \cite{Abb1}, \cite{Crot}). The advantages of the double gap design are clear within the model, i.e. doubling the induced signal while keeping the charge released per gap at the same level than in a single gap. The electrodes are made out of a Bakelite which has lower resistivity than glass and the Bakelite resistivity is controlled by adding water vapours to the gas mixture \cite{NOICsm}.
\par
\begin{figure}[h]
\centering
\includegraphics[width=.6\textwidth]{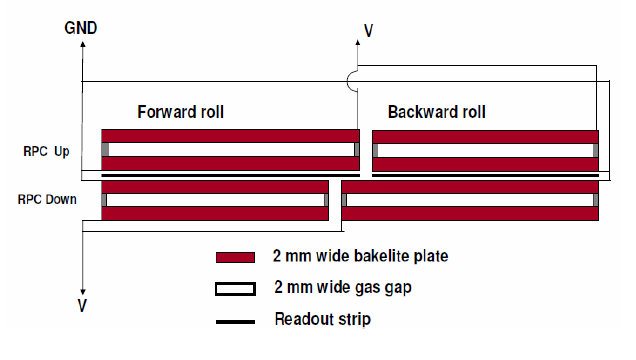}
\caption{Schematic layout of a RPC chamber composed of two double gaps.}
\label{fig:doublegap}
\end{figure}

 Each chamber consists of two or three adjacent double-gaps (Figure \ref{fig:doublegap}) held together by aluminium profiles. There are 1020 double-gaps in the barrel and 1296 in the end-caps. Each double-gap has up to 96 read-out strips. There are more than $10^5$ read-out channels in total for the whole RPC subsystem. The chambers in the barrel have rectangular geometry and the width of the strips varies from $2.3$ cm for the innermost muon station to $4.1$ cm for the outermost muon station. Chambers in the end-caps have trapezoidal geometry and the width of the strips varies from $1.7$ cm at the lowest radii to roughly $3.6$ cm at the highest radii.

The RPCs have been used as dedicated muon trigger detectors. Due to the time resolution of about $1.8$ {ns}, proper association between the RPC trigger information and the LHC bunch crossing has been ensured \cite{NOICsm}.

\section{The CMS RPC simulation}
CMS experiment should be able to distinguish the detector related effects from the real physical processes. The CMS experiment studies different physical processes, detecting the particles originating from proton-proton or lead-lead collisions. A crucial part of the data analysis is the successful factorization of the detector related effects from the real physical processes, resulting in the collisions. For example one needs to know very well the geometrical acceptance and efficiency of the detector and the trigger system in order to calculate the cross section for a given physical process. The CMS detector performance has been very well studied in details and has been implemented in the CMS simulation, thus simulation has been used to account for detector related effects, during data analysis.  A dedicated Monte Carlo simulation of the CMS detector is a crucial part of the data analysis and the simulation result enters in the final physical result of the CMS collaboration. The correct simulation of the detector response is used also for the study of the upgrade of the CMS RPC system.
In the past decades several RPC simulation techniques are proposed by different authors \cite{Ava}, \cite{RpcSim}, \cite{RpcSimNeu}, \cite{Lipp}, \cite{DiePos}. Most of them are focused on the simulation of the RPC properties ,,ab initio``, i.e. obtaining detector characteristics (charge spectra, efficiency, timing, etc.) from detector parameters (e.g. Townsend coefficient, gas gap width, number of gaps, etc.). These techniques are very important and useful for detector R{\&}D. They can be used to simulate different RPC types and to optimize detector design in terms of gap width, gas composition, strip parameters. Even an approach for modeling of the RPC behavior, based on artificial neural networks is developed \cite{Cola}. The method predicts the behavior of the anode charge as a function of the environmental temperature and pressure.
All of these techniques are good for describing the parameters of a given RPC type.
Our task is to simulate 2316 individual detectors, all together with the detector electronics. All of these detectors are of one type in terms of Townsend coefficients, efficiency and front-end electronics. The detectors are working for almost a decade and the important detector  parameters have been measured and well understood. To take advantage of our knowledge we adopted data driven parametrized simulation model. Collision data and cosmic data collected in 2011 have been used for estimation of the RPC efficiency and noise simulation parameters. The time resolution and cluster size distribution have been estimated from experimental data collected on cosmic rays \cite{NOICsm}.

The following parameters, taken from a dedicated database have been used in the default digitization algorithm:

\begin{itemize}
\item efficiency for each readout strip;
\item cluster size distributions;
\item detector time resolution.
\item noise rate for each readout strip;
\end{itemize}

In order to tune the MC simulation to the experimental data it is very important to update regularly the parameters used for the RPC response simulation. Presently, three sets of parameter values with different efficiency and noise rate parameters are used:
\begin{itemize}
\item default values : RPC efficiency $95$\%; strip noise $0.05$ {Hz/cm$^2$}; time resolution $1.42 $ ns; cluster size distribution estimated from experimental data collected by cosmic ray telescope \cite{Abb};
\item parameters estimated using cosmic-ray data;
\item parameters estimated from proton-proton collisions data.
\end{itemize}

\subsection{The simulation chain}
\par CMSSW is a dedicated software framework for CMS event reconstruction, Monte Carlo (MC) simulation and data analysis. Simulation of the RPC response is an integral part of the entire CMSSW simulation chain which is used for the physical CMS analysis.  The simulation chain in CMSSW has been organized as follows:

\begin{itemize}
\item event generation -- events have been generated by physics events generators, interfaced by CMSSW. In the particular case Pythia event generator \cite{PYT} has been used in order to generate Drell--Yan events with two muons in the final state \cite{DlY};
\item event simulation -- CMSSW interface to GEANT4\cite{GEA} has been used to simulate the passage of particles through the detector. The whole CMS detector has been described precisely in the simulation, taking into account the mechanical design, material budget and magnetic field. The RPCs have been described by their components i.e. aluminium housing, Bakelite plates, active gas volume, filled with the ,,standard`` gas mixture \cite{NOICsm} and readout strips. In the particular case of RPC simulation, the result of the simulation step is a simulated hit in the active detector volume;
\item digitization -- in response to each simulated hit, one or few adjacent strips have been fired with a given probability. Timing information has been used to assign the detector response to a given bunch crossing. The intrinsic detector noise has been simulated on the same step;
\item trigger -- Level 1(L1) trigger has been emulated and High Level Trigger(HLT) algorithms have been executed on this step;
\item event reconstruction -- the reconstruction algorithms have been performed.
\end{itemize}

\subsection{The simulation parameters}

\begin{itemize}
\item{Efficiency} -- the RPC efficiency has been estimated employing the segment extrapolation method\cite{CAR}.
DT/CSC reconstructed track segments have been extrapolated to the RPC strip planes and used to look for RPC hits in the surroundings of the extrapolated point. Efficiency has been computed as the ratio between the number of observed RPC hits and the expected number of hits from segment extrapolation. The average chamber efficiency computed with this method has been assigned to all the strips of the same chamber. These values obtained from experimental data have been used for the RPC efficiency parametrization. During the simulation, the decision whether the strip should be fired or not has been taken according to the assigned strip efficiency.

\item{Cluster size} -- the number of simultaneously fired adjacent strips (cluster size) has been calculated using an empirical cluster size distribution for each chamber. The RPC intrinsic efficiency and cluster size are assumed to be independent from the muon momentum in our simulation model. For our studies only high momenta muons (above few tens of GeV) have been taken into account. In a good approximation they behave as minimum ionization particles in our detectors \cite{PDG}. For even higher momenta muons (in TeV range) the radiative effects are treated on the simulation step by GEANT. According to the experimental data \cite{NOICsm}, the clusters size has very weak dependence from the muon impact angle for angles smaller then 45$^\circ$ and thus that dependence is not implemented in the simulation. However, detailed analysis of the experimental data shows that the cluster size depends on the muon impact point on the strip. The dependence is parametrized in the simulations and the parameters are obtained from the experimental data. The strip has been formally subdivided in five slices along its width, and the impact point of the muon has been used to find the fired slice coordinates. For each slice we have a different cluster size distribution and the random number, corresponding to the cluster size, is generated from the distribution. All the distributions have been obtained from the experimental data, using an algorithm explained in \cite{Colaleo}. An example of such a distribution is presented on fig. \ref{fig:CLS}.

\item{Timing} -- the read-out of the CMS detector is performed every $25$ ns. The true signals correspond to the signals within a $25$ time window, synchronized with the LHC bunch crossing separation, so called central bunch crossing.  The information from six consecutive time windows, two before and three after the central bunch crossing, has been collected to form an event. For proper bunch crossing assignment the time of flight of the particle, the signal propagation speed on the strip and the delay of the signal along the cables for each chamber have been taken into account. The time of flight depends on the momentum of the muon and has been calculated for each simulated particle. The signal propagation time along the strip depends on the hit position on the strip. It has been simulated assuming the signal propagation speed equals to $66$\% of the speed of light.  The value assumed for the signal propagation speed is consistent with the average shift of the signal arrival time observed in data samples taken at different positions along the read-out strips for a set of chambers \cite{Bru}. The uncertainty of the signal propagation speed is not an issue in the simulation, because the time is measured in bunch-crossings, i.e. time quanta is $25$ ns. The intrinsic time resolution of the chambers altogether with the attached front-end electronics is less than $2$ ns for all the chambers and the time resolution has been simulated as a Gaussian with a mean value 0 ns and sigma $1.42$ ns. The different channel-to-channel delays in a chamber are far below this value. A constant time offset for every chamber is added to account for different cable lengths, muon time of flight etc. The RPC signal synchronization is described in \cite{Bunk}.

\item{Noise} -- there are two possible types of parametrization used for intrinsic noise simulation. A parametrization where all the strips have a default value of $0.05$ Hz/cm$^2$ was used in the first half of the 2011. A realistic noise parametrization has been used afterwards. The intrinsic RPC noise has been measured during cosmic runs \cite{NOICsm} and used to model the MC response. The estimation of the simulated noise rate for a given strip in a certain event has been based on the Poisson distribution with mean $\nu$:
$$
\nu = N_j S_{str} t
$$
where $N_j$ is the measured strip noise, $S_{str}$ is the strip area and the total simulated time is $t=n_{bx} . 25 ns$, where $n_{bx}$ is the number of simulated bunch crossings.

\par The simulated noise signals are uniformly distributed in time windows around the central bunch crossing, including the central one.
\end{itemize}

\section{Validation}
\par The input to the digitization is efficiency, cluster size distribution, time resolution and noise rate, normalized by the area. The output of the digitization is a digi. The digi in the RPC case is a pair of two numbers. The first number correspond to the read out channel of the front-end electronics connected to the fired strip and the second number corresponds to the time when the strip is fired, measured  in bunch-crossing units (time quanta is $25$ ns). The digi is used in the CMSSW framework for the consecutive Level 1 trigger emulation and on the reconstruction step. It is difficult to compare digis to the input parameters and in order to validate the whole simulation and the digitization in particular a dedicated analysis has been performed. In the analysis the output of the reconstruction of the simulated events has been used to estimate efficiency from MC , cluster size distribution from MC and timing from MC. The results have been compared to the input parameters or to the real experimental date in order to estimate the goodness of the MC simulation and correctness of the MC model.

\subsection{Validation of the goodness of the simulation}

\par
\begin{figure}[h]
\begin{minipage}{0.49\textwidth}
\includegraphics[width=.9\textwidth]{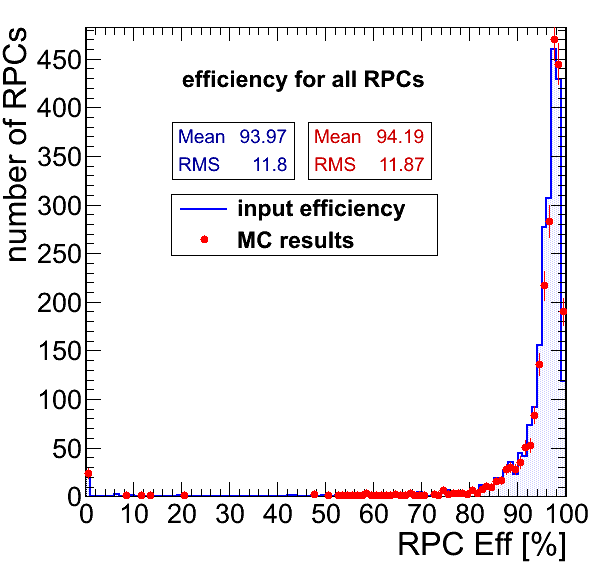}
\end{minipage}
\begin{minipage}{0.49\textwidth}
\includegraphics[width=.9\textwidth]{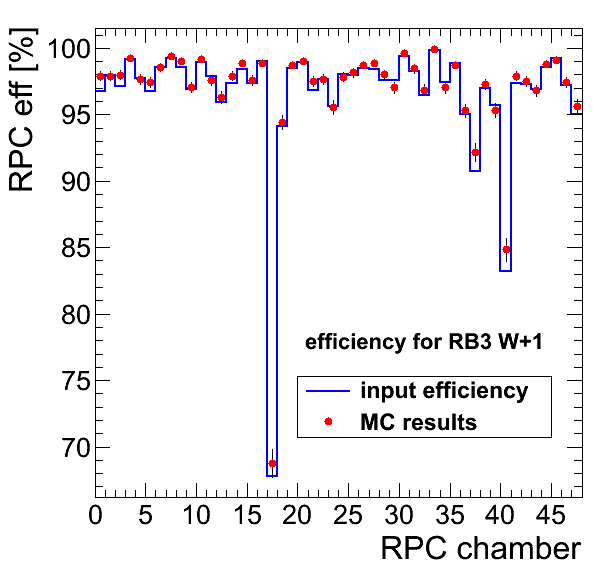}
\end{minipage}
\caption{A comparison between the input parametrization (solid line) and the MC results (dots). Left plot: efficiency distribution for all the RPCs; Right plot: example of the efficiency for all chambers from the fifth layer of one of the Barrel Wheels. Two of the chambers are working in a single gap mode, which is well reproduced by MC.}
\label{fig:Eff}
\end{figure}

 In order to validate the goodness of the simulation, Drell -- Yan events with two muons in the final state with invariant mass greater then $60$ {GeV/c$^2$} have been simulated using the Pythia event generator \cite{PYT}. The generated events have passed the full simulation chain. A dedicated selection procedure choosing muons from Z decays has been applied both on the MC and collision data. For the purposes of validation, the efficiency of a given RPC has been estimated as the ratio between the number of events with at least one fired strip corresponding to a simulated muon hit and the total number of simulated muons crossing the chamber. The comparison between the input efficiency parameters estimated from experimental data and the simulated efficiency is shown on the Figure \ref{fig:Eff}. The chambers working in a single mode have a lower efficiency, which is well reproduced in the MC results.

\par Noise signals in the MC are defined as all the fired strips without a corresponding simulated hit. For the purposes of validation, the average simulated noise for the chamber $\bar{N}_{ch}$ has been calculated as:
$$
\bar{N}_{ch} = \frac{\sum\limits_{n_{ev}}d_{noise}}{S_{ch} T}
$$
where $d_{noise}$ is the number of all noise signals for a given chamber in the event, $S_{ch}$ is the chamber area, $n_{ev}$ is the number of simulated events and $T=n_{ev} . n_{bx} . 25 ns$ is the total simulated time.

\begin{figure}[h]
\begin{minipage}{0.5\textwidth}
\includegraphics[width=.9\textwidth]{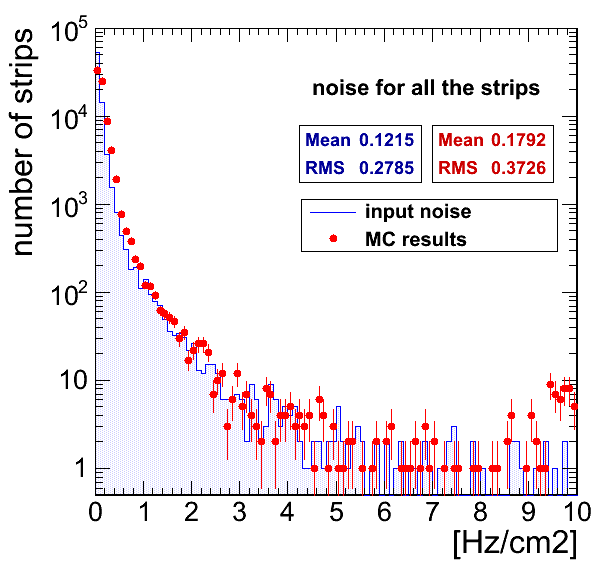}
\end{minipage}
\begin{minipage}{0.5\textwidth}
\includegraphics[width=.9\textwidth]{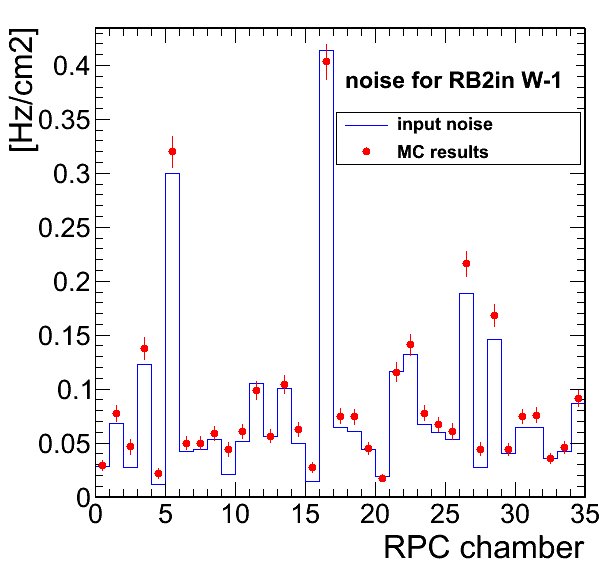}
\end{minipage}
\caption{A comparison between the input parametrization (solid line) and the MC results (dots).
Left plot: Intrinsic noise distribution for all the RPCs strips; Right plot: comparison between the simulated noise and the input noise parameter for all chambers from the third layer of one of the Barrel Wheels.}
\label{fig:NOI}
\end{figure}

The average simulated noise for the strip has been calculated in a similar way, but using the strip area and the number of the noise signals for a given strip. The comparison between the input noise parameters estimated from experimental data and the simulated noise is shown on the Figure \ref{fig:NOI}.
The time resolution, used in the simulation is $1.4$ ns. Thus the timing is always correct, because it is measured in bunch crossing, where one bunch crossing is $25$ ns.

\par
\begin{figure}[h]
\centering
\includegraphics[width=.4\textwidth]{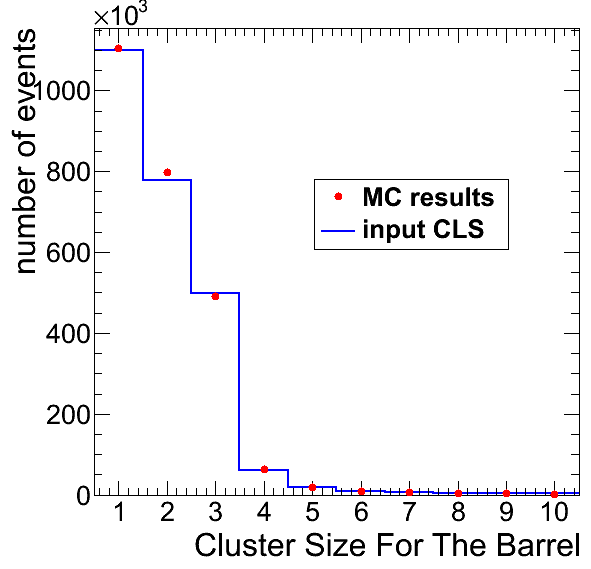}
\includegraphics[width=.4\textwidth]{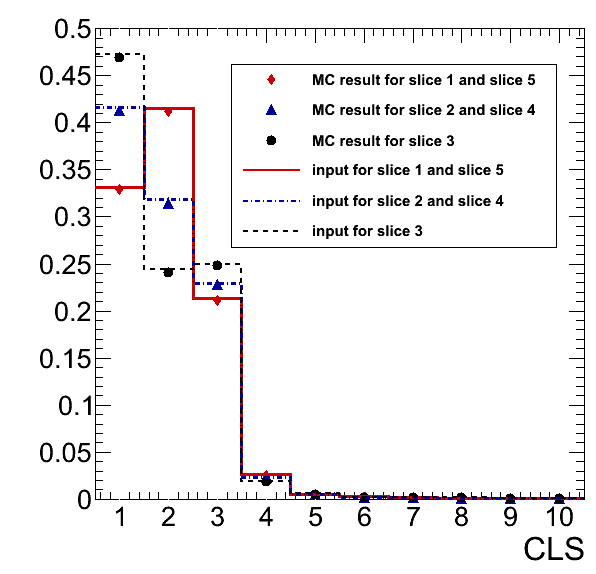}
\caption{Cluster size for muons crossing RPCs. The mark presents the MC result and the line presents the input values. Overall cluster size is presented on the left plot. A comparison between input parameters (cluster size distribution) and MC result for five different muon impact positions (slices) is presented on the right plot.}
\label{fig:CLS}
\end{figure}

 The comparison between cluster size distributions used as simulation input parameters and the MC results is presented on Fig. \ref{fig:CLS} (right) together with overall cluster size distribution (left). The cluster size depends on track impact position, relative to the strip edge. In our parametrization the strip is divided into 5 slices, i.e. the slice corresponds to the muon impact point.

\subsection{RPC system simulation vs experimental data}

\par
\begin{figure}[h]
\begin{minipage}{0.5\textwidth}
\includegraphics[width=.9\textwidth]{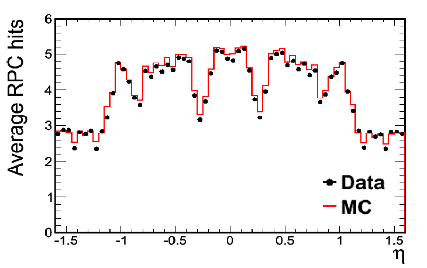}
\end{minipage}
\begin{minipage}{0.5\textwidth}
\includegraphics[width=1.\textwidth]{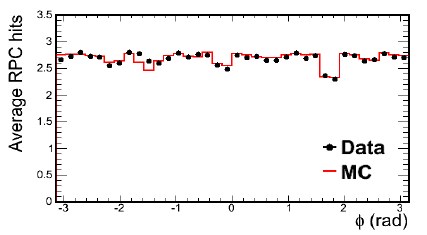}
\end{minipage}
\caption{Left plot: The average number of RPC hits associated to the muon as function of $\eta$ for 
muons with $P_T>20$ GeV/c from $Z$ decay.
Right plot: The average number of RPC hits associated to the muon as function of $\varphi$ for muons with $P_T > 20$ GeV/c from $Z$ decay.  Muons reconstructed both in the tracker and muon system are taken into account in this plot. The dots represent the experimental data and the solid line -- the MC result.
}
\label{fig:MSK}
\end{figure}

 RPCs have been mainly used as dedicated trigger detectors, however, they contribute to the muon reconstruction as well \cite {MSK}. The main goal of the simulation is to reproduce correctly the response of the RPC system, including efficiency, acceptance, misalignment, geometry, etc. To validate the correctness of the overall RPC simulation, the MC simulation has been compared to the real experimental data. The data is taken with the CMS detector during proton-proton collision runs with the energy $\sqrt{s}=7$ TeV at the LHC in 2011. The environmental temperature during data taking is stable. The applied high voltage is corrected for environmental pressure and temperature variations \cite {Const}. The data analysis procedure is described in details in \cite{MSK}. Here, the simulated average number of RPC hits associated to the muon has been compared to the number obtained from experimental data. The number of RPC hits used in the muon reconstruction is related to the number of RPC stations crossed by the muon (that is a function of $\eta$) and of the intrinsic RPC efficiency. The comparison between the average number of reconstructed RPC hits along the muon track for data and MC is shown in Figure \ref{fig:MSK} as a function of $\eta$ and azimuthal angle $\varphi$. The shapes of the plots reflect the geometrical acceptance of the RPC muon system. The simulation model reproduces this feature well.

\section{Conclusion}
\par An algorithm for the simulation of the CMS RPC system has been developed and implemented in the CMS software package CMSSW. The set of simulation parameters has been defined using experimental data taken during cosmic and proton-proton collision runs. The parameter's values habe been regularly updated. A dedicated validation procedure has been developed. The MC and experimental data are in good agreement.

\section{Acknowledgments}
\par We congratulate our colleagues in the CERN accelerator departments for the excellent performance of the LHC machine. We thank the technical and administrative staff at CERN and other CMS institutes, and acknowledge support from BMWF and FWF (Austria); FNRS and FWO (Belgium); CNPq, CAPES, FAPERJ, and FAPESP (Brazil); MEYS (Bulgaria); CERN; CAS, MoST, and NSFC (China); COLCIENCIAS (Colombia); MSES (Croatia); RPF (Cyprus); MoER, SF0690030s09 and ERDF (Estonia); Academy of Finland, MEC, and HIP (Finland); CEA and CNRS/IN2P3 (France); BMBF, DFG, and HGF (Germany); GSRT (Greece); OTKA and NKTH (Hungary); DAE and DST (India); IPM (Iran); SFI (Ireland); INFN (Italy); NRF and WCU (Korea); LAS (Lithuania); CINVESTAV, CONACYT, SEP, and UASLP-FAI (Mexico); MSI (New Zealand); PAEC (Pakistan); MSHE and NSC (Poland); FCT (Portugal); JINR (Armenia, Belarus, Georgia, Ukraine, Uzbekistan); MON, RosAtom, RAS and RFBR (Russia); MSTD (Serbia); SEIDI and CPAN (Spain); Swiss Funding Agencies (Switzerland); NSC (Taipei); ThEP, IPST and NECTEC (Thailand); TUBITAK and TAEK (Turkey); NASU (Ukraine); STFC (United Kingdom); DOE and NSF (USA).

\end{document}